# Charge scattering and mobility in atomically thin semiconductors

Nan Ma[1] and Debdeep Jena
Department of Electrical Engineering, University of Notre Dame
Notre Dame, IN 46556 USA

*Abstract:*

The electron transport properties of atomically thin semiconductors such as $MoS_2$ have attracted significant recent scrutiny and controversy. In this work, the scattering mechanisms responsible for limiting the mobility of single layer semiconductors are evaluated. The roles of individual scattering rates are tracked as the 2D electron gas density is varied over orders of magnitude at various temperatures. From a comparative study of the individual scattering mechanisms, we conclude that all current reported values of mobilities in atomically thin transition-metal dichalcogenide semiconductors are limited by ionized impurity scattering. When the charged impurity densities are reduced, remote optical phonon scattering will determine the ceiling of the highest mobilities attainable in these ultrathin materials at room temperature. The intrinsic mobilities will be accessible only in clean suspended layers, as is also the case for graphene. Based on the study, we identify the best choices for surrounding dielectrics that will help attain the highest mobilities.

---

[1] nma@nd.edu

Two-dimensional (2D) layered crystals such as single layers of transition-metal dichalcogenides represent the thinnest possible manifestations of semiconductor materials that exhibit an energy bandgap. For example, a single-layer (SL) $MoS_2$ is ~0.6 nm thick, and exhibits an energy bandgap of ~1.8 eV [1]. Such semiconductor layers differ fundamentally from ultrathin heterostructure quantum wells, or thin membranes carved out of three-dimensional (3D) semiconductor materials because there are in principle no broken bonds, and no roughness over the 2D plane. In heterostructure quantum wells, the electron mobility suffers from variations in the quantum-well thickness. A classic 'sixth-power law' due to Sakaki et al. [2] shows that since the quantum-mechanical energy eigenvalues in a heterostructure quantum well of thickness $L$ go as $\varepsilon \sim 1/L^2$, variations in thickness $\Delta L$ lead to perturbations of the energy $\Delta \varepsilon \sim -2\Delta L/L^3$. Since the scattering rate depends on the square of $\Delta \varepsilon$, the roughness-limited mobility degrades as $\mu_R \sim L^6$. When $L$ reduces from ~7 to ~5 nm for example, $\mu_R$ reduced from ~$10^4$ to $10^3$ $cm^2$/Vs in GaAs/AlAs quantum wells at 4.2 K [2]. Though low-temperature mobilities exceeding $10^6$ $cm^2$/Vs have been achieved in such heterostructures by scrupulous cleanliness and design to reduce roughness scattering, the statistical variations in the quantum well thickness during the epitaxial growth process pose a fundamental limit to electron mobility.

Due to the absence of intrinsic roughness in atomically thin semiconductors, the expectation is that higher mobilities should, in principle, be attainable. However, recent measurements in $MoS_2$ and similar semiconductors [3-5] exhibit rather low mobilities in single layers, which are, in fact, lower than in their multilayer counterparts. Many-particle transport effects can appear in transition-metal dichalcogenides under special conditions due to the contribution of highly localized d-orbitals to the conduction and valence band edge eigenstates. Collective effects have been observed in multilayer structures, such as charge-density waves [6,7] and the appearance of superconductivity at extremely high metallic carrier densities [8] under extreme conditions. We do not discuss such collective phenomena here, and focus the work on single-particle transport in single-layer $MoS_2$; the only many-particle effect included is free-carrier screening. In this work, we perform a comprehensive study of the scattering mechanisms that limit electron mobility in atomically thin semiconductors. The mobility is calculated in the relaxation-time approximation (RTA) of the Boltzmann Transport Equation. The results shed light on the experimentally achievable electron mobility by designing the surrounding dielectrics and lowering the impurity density. The findings thus offer useful guidelines for future experiments.

With the advent of graphene, it was realized that for ultrathin semiconductors, the dielectric environment plays a crucial role in electron transport. It has now been demonstrated that the dielectric mismatch significantly modifies the Coulomb potentials inside a semiconductor thin layer [9-12]. Electrons in the semiconductor can also remotely excite polar optical phonon modes in the dielectrics [13-19]. Such long-range interactions become stronger as the thickness of the semiconductor layer decreases. Thus one can expect the dielectric environment to significantly affect electron transport properties in SL gapped semiconductors. In this work, we take SL MoS$_2$ as a case study to investigate such effects. The results and conclusions can be extended to other SL gapped semiconductors.

We first study the effect of the dielectric environment on Coulomb scattering of carriers from charged impurities located inside the MoS$_2$ single layer. Figure 1 (a) shows a point charge located at the center ($z_0 = 0$) of a SL MoS$_2$ of thickness $a$. Assuming the surrounding dielectric provides a large energy barrier for confining electrons in the MoS$_2$ membrane, we consider scattering of electrons within the conduction band minima at the *K*-point, i.e., in the ground state. The envelope function of mobile electrons then is: $\psi_{\vec{k}}(\vec{\rho},z) = \chi(z)e^{i\vec{k}\cdot\vec{\rho}}/\sqrt{S}$, where $\chi(z) = \sqrt{2/a}\cos(\pi z/a)$, $S$ is the 2D area, $\vec{k}$ is the in-plane 2D wave vector, and $\vec{\rho}$ is the in-plane location vector of the electron from the point charge. The dielectric mismatch between the MoS$_2$ (relative dielectric constant $\varepsilon_s$) layer and its environment ($\varepsilon_e$) creates an infinite array of image charges at points $z_n = na$, where $n = \pm 1, \pm 2...$ [9,10,20]. The $n^{th}$ point charge has a magnitude of $e\gamma^{|n|}$, where $\gamma = (\varepsilon_s - \varepsilon_e)/(\varepsilon_s + \varepsilon_e)$. These image charges contribute to the net electric potential seen by the electron, which is given by

$$V_{unsc}^{CI}(\rho,z) = \sum_{n=-\infty}^{\infty} \frac{e\gamma^{|n|}}{4\pi\varepsilon_0\varepsilon_s\sqrt{\rho^2 + |z-z_n|^2}} \quad (1)$$

Figure 1 shows the net *unscreened* Coulomb potential contours in the dielectric/MoS$_2$/dielectric system with three different $\varepsilon_e$. The Coulomb interaction is strongly enhanced for low-κ dielectric environment and is damped for the high-κ case.

When a point charge is located inside a 3D semiconductor, its Coulomb potential is lowered by the dielectric constant of the semiconductor host alone. For thin semiconductor layers, the Coulomb potential is determined by the dielectric constants of both the semiconductor itself as well as the surrounding dielectrics. When a high density of mobile carriers is present in the semiconductor, the Coulomb potential is further screened. For atomically thin semiconductors, understanding the dielectric mismatch effect on the free-carrier screening of scattering potentials is necessary. At zero temperature, screening by the 2D electron gas is captured by the Lindhard function [21]:

$$\varepsilon_{2d}(q,0) = 1 + \frac{e^2}{2\varepsilon_0\varepsilon_s q}\Pi(q,0)(\Phi_1 + \Phi_2), \quad (2)$$

where $q$ is the 2D scattering wave vector, $\Pi$ is the polarizability function at zero temperature [22],

$$\Pi(q,0) = \frac{g_s g_v m^*}{2\pi\hbar^2}\left\{1 - \Theta[q-2k_F]\sqrt{1-\left(\frac{2k_F}{q}\right)^2}\right\} \quad (3)$$

where $\Theta[...]$ is the Heaviside unit-step function. The function $\Phi_1$ is the form factor, and $\Phi_2$ is the dielectric mismatch factor, which are defined by the equations [23]

$$\Phi_1 = \int \chi^2(z)dz \int \chi^2(z')\exp(-q|z-z'|)dz' \quad (4)$$

$$\Phi_2 = \frac{2\chi_+\chi_-\exp(-qa)(\varepsilon_e - \varepsilon_s)^2 - (\chi_-^2 + \chi_+^2)(\varepsilon_e^2 - \varepsilon_s^2)}{\exp(qa)(\varepsilon_e + \varepsilon_s)^2 - \exp(-qa)(\varepsilon_e - \varepsilon_s)^2} \quad (5)$$

where $\chi_\pm = \int dz \exp(\pm qz)\chi^2(z)$. The free-carrier screening is taken into account by dividing the unscreened scattering matrix elements by $\varepsilon_{2d}$. Eq. (2) can be re-cast as the Thomas-Fermi formula:

$\varepsilon_{2d} = 1 + q_{TF}^{eff}/q$ in analogy to the case in the absence of dielectric mismatch. Here $q_{TF}^{eff}$ corresponds to the Thomas-Fermi screening wave vector $q_{TF}^0$ without dielectric mismatch. Figure 2 (a) shows the ratio $q_{TF}^{eff}/q_{TF}^0$ that captures the effect of the dielectric mismatch on screening at zero temperature. The 2D electron density is $n_s \sim 10^{12}$ cm$^{-2}$ in this figure. As can be seen, the free-carrier screening is weakened by a high-κ dielectric, and is enhanced in a low-κ case. This dependence is *opposite* to the effect of the dielectric environment on the net *unscreened* Coulomb interaction.

The momentum relaxation rate $(\tau_m)^{-1}$ due to elastic scattering mechanisms is evaluated using Fermi's golden rule in the form

$$\frac{1}{\tau_m} = \frac{2\pi}{\hbar} \int \frac{d^2 k'}{(2\pi)^2} \frac{|M_{kk'}|^2}{\varepsilon_{2d}^2} (1 - \cos\theta) \delta(E_k - E_{k'}), \qquad (6)$$

where $M_{kk'}$ is the matrix element for scattering from state $k$ to $k'$, $\theta$ is the scattering angle, $E_k$ and $E_{k'}$ are the electron energies for states $k$ and $k'$, respectively. For the charged impurity scattering momentum relaxation rate $(\tau_m^c)^{-1}$, the scattering matrix element is evaluated as

$$M_{kk'} = \frac{e^2}{2\varepsilon_0 \varepsilon_s S} \frac{1}{q} \times 4 \left\{ \frac{\gamma}{\exp(qa) - \gamma} \frac{4\pi^2 \sinh(\frac{qa}{2})}{4\pi^2(qa) + (qa)^3} + \frac{2\left[1 - \exp(-\frac{qa}{2})\right]\pi^2 + (qa)^2}{4\pi^2(qa) + (qa)^3} \right\} \qquad (7)$$

Fig. 2 (b) shows $(\tau_m^c)^{-1}$ with the impurity density of $N_I \sim 10^{12}$ cm$^{-2}$. $\varepsilon_e$ and $n_s$ are varied over 2 orders of magnitude to map out the parameter space. Evidently, $(\tau_m^c)^{-1}$ still reduces monotonically with increasing $\varepsilon_e$ because the weakening of the unscreened Coulomb potential wins out.

The reduction of $(\tau_m^c)^{-1}$ for a high-κ environment is much enhanced for high $n_s$, as indicated in Fig. 2 (b). When $\varepsilon_e$ varies from 1 to 100, $(\tau_m^c)^{-1}$ decreases ~1.4 times for $n_s \sim 10^{11}$ cm$^{-2}$, and ~2.6 times for $n_s \sim 10^{13}$ cm$^{-2}$. From the perspective of screening, notice from Fig. 2 (a) that in a low-κ environment, $q_{TF}^{eff}$ is higher for small angle scattering events. This means the smaller the scattering angle, the stronger is the screening. Thus screening *favors* randomizing the electron momentum. A high-κ environment reverses this process: small angle scattering events are weakly screened, and thus such scattering events are favored. Thus, as $\varepsilon_e$ increases, the electron transport become more *directional*. Though $q_{TF}^{eff}$ decreases, the net screening efficiency increases. These tendencies are enhanced as $n_s$ increases. From the scattering potential point of view, a higher $n_s$ leads to a larger Fermi wave vector $k_F$. As shown schematically in the inset of Fig. 2 (a), the same $q = |\mathbf{k}_i - \mathbf{k}_f|$ with high $n_s$ corresponds to a smaller scattering angle than a lower $n_s$ case, leading to a reduced $(\tau_m^c)^{-1}$. This effect on the Coulomb scattering matrix element is multiplied by the dielectric mismatch factor, thus a high-$n_s$ system shows stronger $\varepsilon_e$-dependence at zero temperature.

For finite temperatures, following Maldague, the polarizability function is [22,24,25]:

$$\Pi(q,T,E_F) = \int_0^\infty \frac{\Pi(q,0)}{4k_B T \cosh^2[(E_F - E)/2k_B T]} dE, \qquad (8)$$

where $E_F$ is the Fermi energy, and $k_B$ is the Boltzmann constant. Figure 3 (a) shows the calculated temperature-dependent polarizability normalized to the zero-temperature value at different $n_s$. The electron gas is less polarizable at higher temperatures and lower $n_s$. Polarizability is caused by the spatial re-distribution of the electron gas induced by the Coulomb potential, thus it is proportional to $n_s$. As temperature increases, the thermal energy randomizes the electron momenta, accelerating the transition of the electron system back into an equilibrium distribution, consequently weakening the polarization. The decrease of polarizability reduces the free-carrier screening. Figure 3 (b) shows the temperature-dependent Coulomb-scattering-limited mobility ($\mu_{imp}$) at two different $n_s$. The dielectric mismatch effect is more significant for low $n_s$, because of the fast decrease of the polarizability with increasing temperature. For high $n_s$ on the other hand, the dielectric mismatch effect is not as drastic.

The shape of the temperature-dependent $\mu_{imp}$ curve is highly dependent on the polarizability and $n_s$. Consequently, if the electron transport is dominated by impurity scattering, one can infer $n_s$ from the *shape* of the temperature dependence of the electron mobility.

Much interest exists in using atomically thin semiconductors as possible channel materials for electronic devices, in which such layers are in close proximity to dielectrics. To that end, we investigate both the intrinsic and extrinsic phonon scattering in SL MoS$_2$. Kaasbjerg et al. [26] have predicted the theoretical intrinsic phonon-limited mobility ($\mu_{i-ph}$) of SL MoS$_2$ from first principles using a density-functional-based approach. They estimated a room temperature upper-limit for the experimentally achievable mobility of ~410 cm$^2$/Vs, which weakly depended on $n_s$. Their estimate did not include the effects of free-carrier screening and dielectric mismatch. In light of the strong effect of these factors on the Coulomb scattering, we evaluate $\mu_{i-ph}$ in MoS$_2$ in the Boltzmann transport formalism with the modified free-carrier screening. The material parameters for SL MoS$_2$ were obtained from Ref. [27]. The momentum relaxation rate due to quasi-elastic scattering by acoustic phonon is given by

$$\frac{1}{\tau_m^{ac}} = \frac{\Xi_{ac}^2 k_B T m^*}{2\pi \hbar^3 \rho_s v_s^2} \int_{-\pi}^{\pi} \frac{(1-\cos\theta)d\theta}{\varepsilon_{2d}^2} \qquad (9)$$

where $\rho_s$ is the areal mass density of SL MoS$_2$, $v_s$ is the sound velocity, and $\Xi_{ac}$ is the acoustic deformation potential. For inelastic electron-optical phonon interactions, the momentum relaxation rate in the RTA is obtained by summing the emission and absorption processes,

$$\frac{1}{\tau_m^{op}} = \frac{\Theta\left[E_k - \hbar\omega_{op}^v\right]}{\tau_{op}^+} + \frac{1}{\tau_{op}^-}, \qquad (10)$$

where $\omega_{op}^v$ is the frequency of the $v$th optical-phonon mode. The momentum relaxation rates with superscript '+' and '-' are associated with phonon emission and absorption, respectively. For optical deformation potentials (ODP) [26],

$$\frac{1}{\tau_{0-ODP}^{\pm}} = \frac{D_0^2 m^*(N_q + \frac{1}{2} \pm \frac{1}{2})}{4\pi\hbar^2 \rho_s \omega} \int_{-\pi}^{\pi} \frac{(1-\frac{k'}{k}\cos\theta)d\theta}{\varepsilon_{2d}^2}, \text{ and} \quad (11)$$

$$\frac{1}{\tau_{1-ODP}^{\pm}} = \frac{D_1^2 m^*(N_q + \frac{1}{2} \pm \frac{1}{2})}{4\pi\hbar^2 \rho_s \omega} \int_{-\pi}^{\pi} \frac{q^2(1-\frac{k'}{k}\cos\theta)d\theta}{\varepsilon_{2d}^2}, \quad (12)$$

where $D$ is the optical deformation potential, $N_q = 1/[\exp(\hbar\omega/k_BT)-1]$ is the Bose-Einstein distribution for optical phonons of energy $\hbar\omega$, and the subscript 0 and 1 denote the zero- and first-order ODP, respectively.

The scattering rate by polar optical (LO) phonons is given by the Fröhlich interaction [28],

$$\frac{1}{\tau_{LO}^{\pm}} = \frac{e^2\omega m^*}{8\pi\hbar^2} \frac{1}{\varepsilon_0}(\frac{1}{\varepsilon_\infty} - \frac{1}{\varepsilon_s})(N_q + \frac{1}{2} \pm \frac{1}{2})\int_{-\pi}^{\pi} \frac{1}{q}\Phi_1 \frac{(1-\frac{k'}{k}\cos\theta)d\theta}{\varepsilon_{2d}^2} \quad (13)$$

where $\varepsilon_\infty$ is the high frequency relative dielectric constant, and $\Phi_1$ is the form factor defined by Eq. (4).

Figure 4 (a) shows the $n_s$-dependent screened $\mu_{i-ph}$ at room temperature. For comparison, the unscreened $\mu_{i-ph}$ is also shown as a reference by the blue line. The unscreened values remain effectively constant (~380 cm$^2$/Vs) over the range of $n_s$ of interest ($10^{11}$~$10^{13}$ cm$^{-2}$). This is in agreement with the previous predictions (320~410 cm$^2$/Vs) [26,29]. However, the *screened* $\mu_{i-ph}$ increases sharply with increasing $n_s$. As can be seen in Fig. 4 (a), introducing a high-κ dielectric leads to a reduction of $\mu_{i-ph}$; the highest values of $\mu_{i-ph}$ reduce from 3100 to 1500 cm$^2$/Vs as $\varepsilon_e$ increases from ~7.6 to ~20. The strong dependence of $\mu_{i-ph}$ on the dielectric environment is entirely due to the dielectric-mismatch effect on free-carrier screening, since the unscreened phonon scattering matrix element is not affected by $\varepsilon_e$. Over the entire range of $n_s$, longitudinal optical phonon scattering is dominant. This finding is different from previous works on multi-layer MoS$_2$ transport where the room temperature $\mu_{i-ph}$ was determined by homopolar phonon scattering [30-32].

We have used the static dielectric function for calculating the screened interactions due to different modes of phonons in the limit $\omega \to 0$. Scattering mechanisms via long-range Coulomb interactions, such as charged impurities, polar optical phonons, and piezoelectric acoustic phonons, can be effectively screened by free carriers. However, free carriers *may* not respond to rapidly changing scattering potentials originating from short-range interactions. There are arguments about to what extent the short-range deformation potentials induced by acoustic (ADP) and optical phonons (ODP) are screened by free carriers. Boguslawski and Mycielski [33] argue that in a single-valley conduction band, the deformation potentials (both ADP and ODP) are screened in the same way as the macroscopic (long-range) phonon potentials. But for multi-valleys semiconductors (Ge), only the longitudinal acoustic (LA) mode of the ADP can be effectively screened by free carriers. The free-carrier screening of the transverse acoustic (TA) mode ADP, and ODP can to a good approximation be neglected. [34]. In SL MoS$_2$, Kaasbjerg et. al. [27] have argued that the LA mode of ADP can be treated as screened by the long-wavelength dielectric function, while the screening of TA mode ADP by free carriers can be neglected.

Fig 4 (b) highlights the effect of the partially screened electron-phonon interaction compared to the fully screened version in Fig 4(a). For the plot in Fig 4 (b), we have screened the polar optical and LA phonon scattering as in Fig. 4 (a), and leave the TA and ODP interactions unscreened. The highest $\mu_{i-ph}$ reached by free carrier screening effect is reduced to ~750 cm$^2$/Vs by not screening the DP modes. The mobility is dominated by the polar optical phonon interaction at low carrier density and by TA and ODP at moderate and high density. The scattering of electrons due to piezoelectric phonons is not considered because it is relevant only at very low temperatures and there are still uncertainties in the piezoelectric coefficients of SL MoS$_2$ [27, 35].

In both cases, the calculated room temperature $\mu_{i-ph}$ are much higher than reported experimental values, implying that there is still a large room for improvement of mobilities in atomically thin semiconductors. For the rest of this work, we use the fully screened intrinsic phonon scattering as shown in Fig. 4 (a). To pinpoint the most severe scattering mechanisms limiting the mobility in current samples, we discuss an *extrinsic* phonon scattering mechanism at play in these materials, again motivated by similar processes in graphene.

Electrons in semiconductor nanoscale membranes can excite phonons in the surrounding dielectrics via long-range Coulomb interactions, if the dielectrics support polar vibrational modes. Such 'remote phonon' or 'surface-optical' (SO) phonon scattering has been investigated recently for graphene and found to be far from negligible [15-17]. SO phonon scattering can severely degrade electron mobility; however this process has not been studied systematically in atomically thin semiconductors. The electron-SO phonon interaction Hamiltonian is [15,17,18]:

$$H_{e-SO} = eF_v \sum_q [\frac{e^{-qz}}{\sqrt{q}} (e^{i\vec{q}\cdot\vec{\rho}} a_q^{v+} + e^{-i\vec{q}\cdot\vec{\rho}} a_q^v)], \quad (14)$$

where $a_q^{v+}$ ($a_q^v$) represents the creation (annihilation) operator for the $v$th SO phonon mode. Neglecting the dielectric response of the atomically thin MoS$_2$ layer in lieu of the surrounding media, the electron-SO phonon coupling parameter $F_v$ is:

$$F_v^2 = \frac{\hbar \omega_{SO}^v}{2S\varepsilon_0} \left( \frac{1}{\varepsilon_{ox}^\infty + \varepsilon_{ox'}^\infty} - \frac{1}{\varepsilon_{ox}^0 + \varepsilon_{ox'}^\infty} \right), \quad (15)$$

where $\varepsilon_{ox}^\infty$ ($\varepsilon_{ox}^0$) is the high (low) frequency dielectric constant of the dielectric hosting the SO phonon, and $\varepsilon_{ox'}^\infty$ is the high frequency dielectric constant from the dielectric on the other side of the membrane. The frequency of the SO phonon $\omega_{SO}^v$ is [17,36]

$$\omega_{SO}^v = \omega_{TO}^v \left( \frac{\varepsilon_{ox}^0 + \varepsilon_{ox'}^\infty}{\varepsilon_{ox}^\infty + \varepsilon_{ox'}^\infty} \right)^{1/2}, \quad (16)$$

where $\omega_{TO}^v$ is the $v$th bulk transverse optical-phonon frequency in the dielectric. The scattering rate due to SO phonon is then given by

$$\frac{1}{\tau_{SO}^\pm} = \frac{32\pi^3 e^2 F_v^2 m^* S}{\hbar^3 a^2} (N_q + \frac{1}{2} \pm \frac{1}{2}) \int_{-\pi}^{\pi} \frac{1}{q} \frac{\sinh^2(\frac{aq}{2})}{(4\pi^2 q + a^2 q^3)^2} \frac{(1 - \frac{k'}{k}\cos\theta)d\theta}{\varepsilon_{2d}^2} \quad (17)$$

Table I summarizes the parameters for some commonly used dielectrics.

Figure 5 shows the room-temperature electron mobility for various dielectric environments for two representative temperatures, 100 K and 300 K. $N_I$ and $n_s$ are both ~$10^{13}$ cm$^{-2}$. The solid lines show the net mobility by combining the scattering from charged impurities, intrinsic and SO phonons, whereas the dashed lines show the cases *neglecting* the SO phonons. When SO phonon scattering is absent, the electron mobility is limited almost entirely by $\mu_{imp}$, which increases with $\varepsilon_e$ due to the reduction of Coulomb scattering by dielectric screening. The addition of the SO phonon scattering does not change things much at 100 K except for the highest $\varepsilon_e$ case (HfO$_2$/ZrO$_2$). But it drastically reduces the electron mobility at room temperature, as is evident in Fig. 5. For instance, neglecting SO phonon scattering, one may expect that by using HfO$_2$/ZrO$_2$ as the dielectrics instead of SiO$_2$/air, the RT mobility $\mu_{imp}$ should improve from ~45 to ~80 cm$^2$/Vs. However, when the SO phonon scattering is in action, the mobility in HfO$_2$/MoS$_2$/ZrO$_2$ structure is actually degraded to ~25 cm$^2$/Vs, even lower than the SiO$_2$/air case. Thus, SL MoS$_2$ layers suffer from enhanced SO phonon scattering if they are in close proximity to high-κ dielectrics that allow low-energy polar vibrational modes.

To calibrate our calculations, we study the temperature-dependent electron mobility for SL MoS$_2$ embedded between SiO$_2$ and HfO$_2$, and compare the calculations with reported experimental results. This structure is often used in top-gated MoS$_2$ field effect transistors (FETs), thus understanding the transport in it provides a pathway to understand the device characteristics. In Fig. 6 (a), the blue curves indicate calculated values of $\mu_{imp}$ with different $N_I$, and the red line shows the SO phonon scattering limited mobility ($\mu_{SO}$), with $n_s$ ~$10^{13}$ cm$^{-2}$. The temperature-dependent $\mu_{SO}$ of each SO phonon mode follows the Arrhenius rule: $\mu_{SO} \propto \exp(\hbar\omega_0/k_B T)$, and the net $\mu_{SO}$ is dominated by the softest phonon mode with the lowest energy. The black curves indicate the net mobilities considering all scattering mechanisms discussed in this work. The open squares are the experimental results measured by Hall effect on SL MoS$_2$ FETs from Ref. [4]. $N_I$ and $n_s$ necessary to fit the data are indicated in Fig. 6 (a).

At low temperature, the experimental electron mobility in SL MoS$_2$ is entirely limited by $\mu_{imp}$. This is really not unexpected; it took several decades of careful epitaxial growths and ultraclean control to achieve the high mobilities in III-V semiconductors at low temperatures. Based on this study, we predict that the low-temperature mobilities in atomically thin semiconductors can be *significantly* improved by lowering the impurity density. The room-temperature mobility in III-V semiconductors is limited by intrinsic polar-optical phonon scattering. For comparison, we find that for SL MoS$_2$, the room-temperature mobility is considerably degraded by SO phonon scattering, even with $N_I$ as high as $6 \times 10^{12}$ cm$^{-2}$, as shown in Fig. 6. When SO phonon scattering is absent, the room temperature mobility is expected to be ~130 cm$^2$/Vs with $N_I = 6 \times 10^{12}$ cm$^{-2}$, but the measured values are typically lower (~50 cm$^2$/Vs). Consequently, using HfO$_2$ as gate dielectrics can modestly improve $\mu_{imp}$. However the strong SO phonon scattering that comes with HfO$_2$ can severely decrease the high-temperature electron mobility in clean MoS$_2$ with low charged impurity densities.

An important question then is: which dielectric can help in improving the room-temperature electron mobility in SL MoS$_2$? To answer that question, in Fig. 6 (b), we plot the room-temperature (intrinsic+SO) phonon-limited electron mobility ($\mu_{ph}$) in SL MoS$_2$ surrounded by different dielectrics. From the overall trend, $\mu_{ph}$ decreases with increasing $\varepsilon_e$, and suspended SL MoS$_2$ shows the highest potential electron mobility (over 10,000 cm$^2$/Vs). It is worth noting that if the scattering of electrons by intrinsic phonons is only partially screened, as shown in Fig. 4 (b), the highest achievable mobility in SL MoS$_2$ will be an order lower (~1000 cm$^2$/Vs). However these high values are attainable in suspended SL MoS$_2$. Because $\mu_{ph}$ for MoS$_2$ surrounded by high-κ materials is dominated by SO phonon scattering, the values do not vary much. The critical impurity densities ($N_{cr}$) corresponding to $\mu_{imp} = \mu_{ph}$ are shown in Fig. 6 (c). As long as $N_I \geq N_{cr}$, $\mu_{imp}$ completely masks $\mu_{ph}$. When $N_I < N_{cr}$, the electron mobility starts to be dominated by phonons and moves towards the upper-limit. High $\mu_{ph}$ indicates a greater potential for attaining higher electron mobilities. However, it also needs the sample to be highly pure. In high-κ environments that support low-energy polar vibrational modes, there is not as much room for improving the electron mobility as is in low-κ structures. A compromise is seen for AlN and BN based dielectrics, which by virtue of the light atom N allows high-energy optical modes in spite of their polar nature. From Fig. 6 (b) and (c), one can obtain two useful relationships for single-layer MoS$_2$: $\mu_{ph} \sim 35000/\varepsilon_e^{2.2}$ cm$^2$/Vs and $N_{cr} \sim 10^{10}\varepsilon_e^{2.5}$ cm$^{-2}$, with $n_s$ set at a typical on-state carrier density of $10^{13}$ cm$^{-2}$, as shown by dashed lines. These empirical relations should guide the proper choice of dielectrics and the maximum allowed impurity densities.

To further illustrate the relative importance of SO phonon and charged impurity scattering in SL MoS$_2$, we vary $N_I$ and $n_s$ in different dielectric environments and check the changing trends of electron mobilities at room temperature. Figure 7 (a) shows the net electron mobilities in SL MoS$_2$ as a function of $N_I$ with $n_s=10^{13}$ cm$^{-2}$. Figure 7 (b) and (c) show the electron mobility as a function of $n_s$ for $N_I$ =$10^{11}$ and $10^{13}$ cm$^{-2}$. The electron mobility is weakly dependent on the dielectric environment at high $N_I$ (>$10^{13}$ cm$^{-2}$), as shown in the dashed box in the bottom right corner of Fig. 7 (a). Within this window, high-κ dielectrics can improve the mobility, but only very nominally because the unscreened mobilities are already quite low. When $N_I$ is lowered below ~$10^{12}$ cm$^{-2}$, a low-κ environment shows higher electron mobility. For most of the dielectric environments, when $N_I>10^{12}$ cm$^{-2}$, the mobility fits to the following empirical impurity-scattering-dominated relationship: $\mu \approx 4200/[N_I/10^{11}cm^{-2}]$ cm$^2$/Vs, as shown by dashed line in Fig. 7 (a). Using this expression, one can estimate $N_I$ from measured electron mobility for high $n_s$. As $n_s$ decreases, electron mobility in different dielectric environments starts to separate from each other, as shown in Fig. 7 (c). In this case, the electron mobilities can fit to the following relationship: $\mu \approx \frac{3500}{N_I/10^{11}cm^{-2}}\left[A(\varepsilon_e)+(\frac{n_s}{10^{13}cm^{-2}})^{1.2}\right]$ cm$^2$/Vs for $n_s<10^{13}cm^{-2}$, shown as dashed lines in Fig. 7 (c). $A(\varepsilon_e)$ is a fitting constant depending on $\varepsilon_e$, and some values are listed in the inset table of Fig. 7 (c). High-κ dielectrics with low energy phonons ($HfO_2$, $ZrO_2$) severely degrade the electron mobility over the entire $N_I$ range because of the dominant effect of SO phonon scattering. Note that the dielectric mismatch effect can be slightly overestimated here since we have assumed the thickness of the dielectric to be infinite [25]. In top-gated FETs, the top dielectric could be very thin. Thus the capability of improving electron mobility by high-κ dielectrics can be even less significant. Since most applications require high mobilities, high $n_s$, and high $\varepsilon_e$ to be present *simultaneously* in the same structure for achieving the highest conductivities, $AlN/Al_2O_3$ or BN/BN encapsulation emerge as the best compromises among the dielectric choices considered here. One can also perceive of dielectric heterostructures, with a few BN layers closest to $MoS_2$ to damp out the SO phonon scattering, followed by higher-κ dielectrics to enhance the gate capacitance for achieving high carrier densities. All this however requires ultraclean $MoS_2$ to start with, with $N_I$ well below $10^{12}$ cm$^{-2}$ to attain the high room-temperature mobilities ~1000 cm$^2$/Vs. The presence of high impurity densities will always mask the intrinsic potential of the materials, and is the most important challenge moving forward.

In conclusion, carrier transport properties in atomically thin semiconductors are found to be highly dependent on the dielectric environment, and on the impurity density. For current 2D crystal materials, electron mobilities are mostly dominated by charged impurity scattering. Remote phonons play a secondary role at high temperature depending on the surrounding dielectrics. The major point is that the mobilities achieved till date are far below the intrinsic potential in these materials. High-κ gate dielectrics can increase the electron mobility only for samples infected with very high impurity densities. Clean samples with low-κ dielectrics show much higher electron mobilities. AlN and BN based dielectrics offer the best compromise if a high mobility and high gate capacitance are simultaneously desired, as is the case in field effect transistors. The truly intrinsic mobility limited by the atomically thin semiconductor itself can only be achieved in ultraclean suspended samples, just as is the case for graphene.

Acknowledgements: The authors thank Dr. Andras Kis, Dr. Kristen Kaasbjerg, and Deep Jariwala for useful discussions and for sharing experimental data. The research is supported in part by an NSF

ECCS grant monitored by Dr. Anupama Kaul, AFOSR, and the Center for Low Energy Systems Technology (LEAST), one of the six centers supported by the STARnet phase of the Focus Center Research Program (FCRP), a Semiconductor Research Corporation program sponsored by MARCO and DARPA.

References:

[1] K. F. Mak, C. Lee, J. Hone, J. Shan, and T. F. Heinz, *Atomically Thin MoS$_2$: A New Direct-Gap Semiconductor*, Phys. Rev. Lett. **105**, 136805 (2010).

[2] H. Sakaki, T. Noda, K. Hirakawa, M. Tanaka, and T. Matsusue, *Interface roughness scattering in GaAs/AlAs quantum wells*, Appl. Phys. Lett. **51**, 1934 (1987).

[3] D. Jariwala, V. K. Sangwan, D. J. Late, J. E. Johns, V. P. Dravid, T. J. Marks, L. J. Lauhon and M. C. Hersam, *Band-like transport in high mobility unencapsulated single-layer MoS$_2$ transistors*, Appl. Phys. Lett. **102**, 173107 (2013).

[4] B. Radisavljevic and A. Kis, *Mobility engineering and a metal–insulator transition in monolayer MoS$_2$*, Nature Materials **12**, 815 (2013).

[5] H. Fang, S. Chuang, T.C. Chang, K. Takei, T. Takahashi, and A. Javey, *High-Performance Single Layered WSe$_2$ p-FETs with Chemically Doped Contacts*, Nano, Lett. **12**, 3788 (2012).

[6] J. A. Wilson, F. J. Di Salvo, and S. Mahajan, *Charge-Density Waves in Metallic, Layered, Transition-Metal Dichalcogenides*, Phys. Rev. Lett. **32**, 882 (1974).

[7] A. H. Castro Neto, *Charge Density Wave, Superconductivity, and Anomalous Metallic Behavior in 2D Transition Metal Dichalcogenides*, Phys. Rev. Lett. 86, 4382 (2001).

[8] J. T. Ye, Y. J. Zhang, R. Akashi, M. S. Bahramy, R. Arita, and Y. Iwasa, *Superconducting Dome in a Gate-Tuned Band Insulator*, Science **338**, 1193 (2012).

[9] D. Jena and A. Konar, *Enhancement of Carrier Mobility in Semiconductor Nanostructures by Dielectric Engineering*, Phys. Rev. Lett. **98**, 136805 (2007).

[10] A. Konar, and D. Jena, *Tailoring the carrier mobility of semiconductor nanowires by remote dielectrics*, J. Appl. Phys. **102**, 123705 (2007).

[11] A.K.M. Newaz, Y. S. Puzyrev, B. Wang, S. T. Pantelides, and K. I. Bolotin, *Probing charge scattering mechanisms in suspended graphene by varying its dielectric environment*, Nat. Comm. **3**, 734 (2012).

[12] S.-L. Li, K. Wakabayashi, Y. Xu, S. Nakaharai, K. Komatsu, W.-W. Li, Y.-F. Lin, A. Aparecido-Ferreira, and K. Tsukagoshi, *Thickness-Dependent Interfacial Coulomb Scattering in Atomically Thin Field-Effect Transistors*, Nano Lett. **13**, 3546 (2013).

[13] K. Hess and P. Vogl, *Remote Polar Phonon Scattering in Silicon Inversion Layers*, Solid State Commun. **30**, 807 (1979).

[14] B. T. Moore and D. K. Ferry, *Remote polar phonon scattering in Si inversion layers*, J. Appl. Phys. **51**, 2603 (1980).

[15] A. Konar, T. Fang and D. Jena, *Effect of high-κ gate dielectrics on charge transport in graphene-based field effect transistors*, Phys. Rev. B **82**, 115452 (2010).

[16] J.-H. Chen, C. Jang, S. D. Xiao, M. Ishigami and M. S. Fuhrer, *Intrinsic and extrinsic performance limits of graphene devices on $SiO_2$*, Nat. Nano. **3**, 206 (2008).

[17] S. Fratini and F. Guinea, *Substrate-limited electron dynamics in graphene*, Phys. Rev. B **77**, 195415 (2008).

[18] M. V. Fischetti, D. A. Neumayer and E. A. Cartier, *Effective electron mobility in Si inversion layers in MOS systems with a high- κ insulator: the role of remote phonon scattering*, J. Appl. Phys. **90**, 4587 (2001).

[19] L. Zeng, Z. Xin, S. Chen, G. Du, J. F. Kang, and X. Y. Liu, *Remote phonon and impurity screening effect of substrate and gate dielectric on electron dynamics in single layer $MoS_2$*, Appl. Phys. Lett. **103**, 113505 (2013).

[20] M. Kumagai and T. Takagahara, *Excitonic and nonlinear-optical properties of dielectric quantum-well structures,* Phys. Rev. B **40**, 12359 (1989).

[21] F. Stern, *Calculated Temperature Dependence of Mobility in Silicon Inversion Layers*, Phys. Rev. Lett. **44**, 1469 (1980); *Polarizability of a Two-Dimensional Electron Gas*, Phys. Rev. Lett. **18**, 546 (1967).

[22] K. Hirakawa & H. Sakaki, *Mobility of the two-dimensional electron gas at selectively doped n -type $Al_xGa_{1-x}As/GaAs$ heterojunctions with controlled electron concentrations*, Phys. Rev. B **33**, 8291 (1986).

[23] B. A. Glavin, V. I. Pipa, V. V. Mitin and M. A. Stroscio, *Relaxation of a two-dimensional electron gas in semiconductor thin films at low temperatures: Role of acoustic phonon confinement*, Phys. Rev. B **65**, 205315 (2002).

[24] P. Maldague, *Many-body corrections to the polarizability of the two-dimensional electron gas*, Surf. Sci. **73**, 296 (1978).

[25] Z.-Y. Ong and M. V. Fischetti, *Mobility enhancement and temperature dependence in top-gated single-layer MoS$_2$*, Phys. Rev. B **88**, 165316 (2013).

[26] K. Kaasbjerg, K. S. Thygesen, and K. W. Jacobsen, *Phonon-limited mobility in n-type single-layer MoS$_2$ from first principles*, Phys. Rev. B **85**, 115317 (2012).

[27] K. Kaasbjerg, K. S. Thygesen, and A.-P. Jauho, *Acoustic phonon limited mobility in two-dimensional semiconductors: Deformation potential and piezoelectric scattering in monolayer MoS$_2$ from first principles*, Phys. Rev. B **87**, 235312 (2013).

[28] B. L. Gelmont and M. Shur, *Polar opticalphonon scattering in three and twodimensional electron gases*, J. Appl. Phys. **77**, 657 (1995).

[29] X. Li, J. T. Mullen, Z. Jin, K. M. Borysenko, M. Buongiorno Nardelli, and K. W. Kim, *Intrinsic electrical transport properties of monolayer silicene and MoS$_2$ from first principles*, Phys. Rev. B **87**, 115418 (2013).

[30] S. Kim, A. Konar, W.-S. Hwang, J. H. Lee, J. Lee, J. Yang, C. Jung, H. Kim, J.-B. Yoo, J.-Y. Choi et al., *High-mobility and low-power thin-film transistors based on multilayer MoS$_2$ crystals*, Nat. Comm. **3**, 1011 (2012).

[31] R. Fivaz and E. Mooser, *Electron-Phonon Interaction in Semiconducting Layer Structures*, Phys. Rev. **136**, A833 (1964).

[32] R. Fivaz and E. Mooser, *Mobility of Charge Carriers in Semiconducting Layer Structures*, Phys. Rev. **163**, 743 (1967).

[33] P. Boguslawski and J. Mycielski, *Is the Deformation Potential in Semiconductors Screened by Free Carriers?* J. Phys. C: Solid State Phys. **10**, 2413 (1977).

[34] P. Boguslawski, *Screening of the Deformation Potential by Free Electrons in the Multivally Conduction Band*, J. Phys. C: Solid State Phys. **10**, L417 (1977).

[35] K.-A. N. Duerloo, M. T. Ong, and E. J. Reed, *Intrinsic Piezoelectricity in Two-Dimensional Materials*, J. Phys. Chem. Lett. **3**, 2871 (2012)

[36] S. Q. Wang and G. D. Mahan, *Electron Scattering from Surface Excitations*, Phys. Rev. B **6**, 4517 (1972).

[37] V. Perebeinos and P. Avouris, *Inelastic scattering and current saturation in graphene*, Phys. Rev. B **81**, 195442 (2010).

TABLE I. SO phonon modes for different dielectrics.

|  | SiO$_2$[a] | AlN[a] | BN[b] | Al$_2$O$_3$[a] | HfO$_2$[a] | ZrO$_2$[a] |
|---|---|---|---|---|---|---|
| $\varepsilon_{ox}^{0}$ | 3.9 | 9.14 | 5.09 | 12.53 | 23 | 24 |
| $\varepsilon_{ox}^{\infty}$ | 2.5 | 4.8 | 4.1 | 3.2 | 5.03 | 4 |
| $\omega_{SO}^{1}$ | 55.6 | 81.4 | 93.07 | 48.18 | 12.4 | 16.67 |
| $\omega_{SO}^{2}$ | 138.1 | 88.5 | 179.1 | 71.41 | 48.35 | 57.7 |

[a]Ref. [15]

[b]Ref. [37]

Figure captions:

Fig. 1 (Color) Coulomb potential contours due to an on-center point charge for three different dielectric environments: $\varepsilon_e$ =1, 7.6 (= $\varepsilon_s$ ), 100.

Fig. 2 (Color) Effect of dielectric mismatch on the (a) free-carrier screening and (b) Coulomb momentum relaxation rate at zero temperature. Inset of (a) shows schematically the scattering angle for different electron densities.

Fig. 3 (Color) (a) The normalized polarizability and (b) impurity-limited mobility at different electron density as a function of temperature.

Fig. 4 (Color) Electron mobility in MoS$_2$ due to intrinsic phonon scattering at room temperature with the electron-phonon interaction (a) fully screened and (b) partially screened. The dashed lines show mobilities limited by unscreened phonon modes and the solid lines show the mobilities limited by fully screened modes.

Fig. 5 (Color) Electron mobility as function of environment dielectric constant. Dashed lines show the mobility without considering the SO phonons.

Fig. 6 (Color) (a) Temperature-dependent electron mobility (black lines) in SiO$_2$/MoS$_2$/HfO$_2$ structure. The blue lines indicate $\mu_{imp}$ and the red lines show $\mu_{SO}$. Open squares show experimental results from single-layer MoS$_2$ FETs from Ref. [4]. (b) Room temperature phonon-determined electron mobilities $\mu_{ph}$ and (c) the critical impurity densities $N_{cr}$ corresponding to $\mu_{imp} = \mu_{ph}$ in SL MoS$_2$ surrounded by different dielectrics. Dashed lines show the fitted $\mu_{ph}$ and $N_{cr}$.

Fig. 7 (Color) The room temperature net electron mobilities in SL MoS$_2$ with considering all kinds of scattering mechanisms as a function of (a) $N_I$ with fixed $n_s$ at $10^{13}$ cm$^{-2}$; (b) and (c) $n_s$ with $N_I$ fixing at $10^{11}$ and $10^{13}$ cm$^{-2}$, respectively. Numbers on the curves show the average dielectric constant of the surrounding dielectrics. Dashed lines show the fitted electron mobilities.

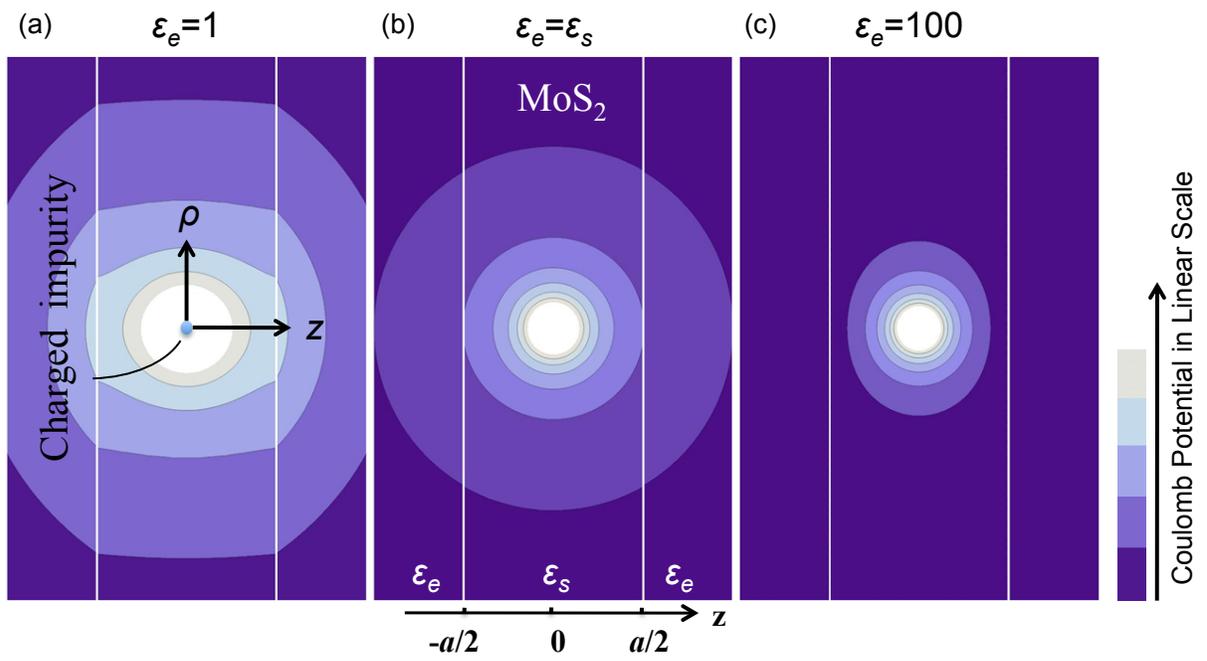

**Figure 1**

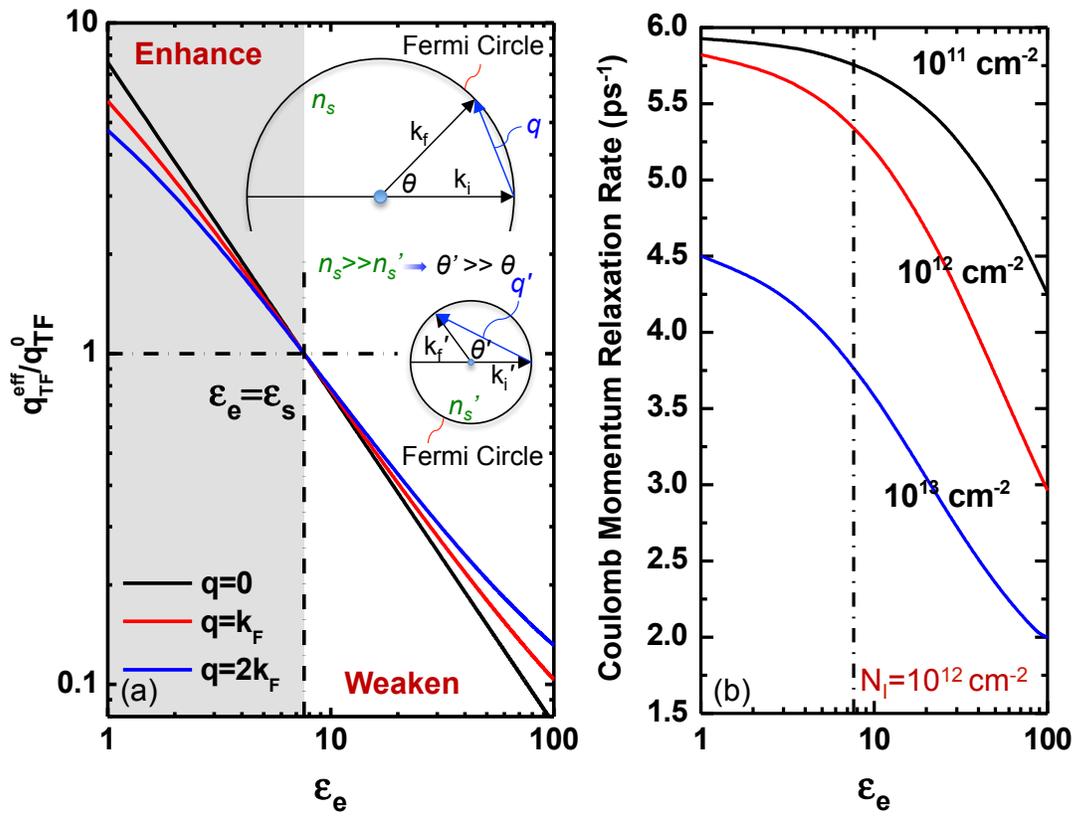

**Figure 2**

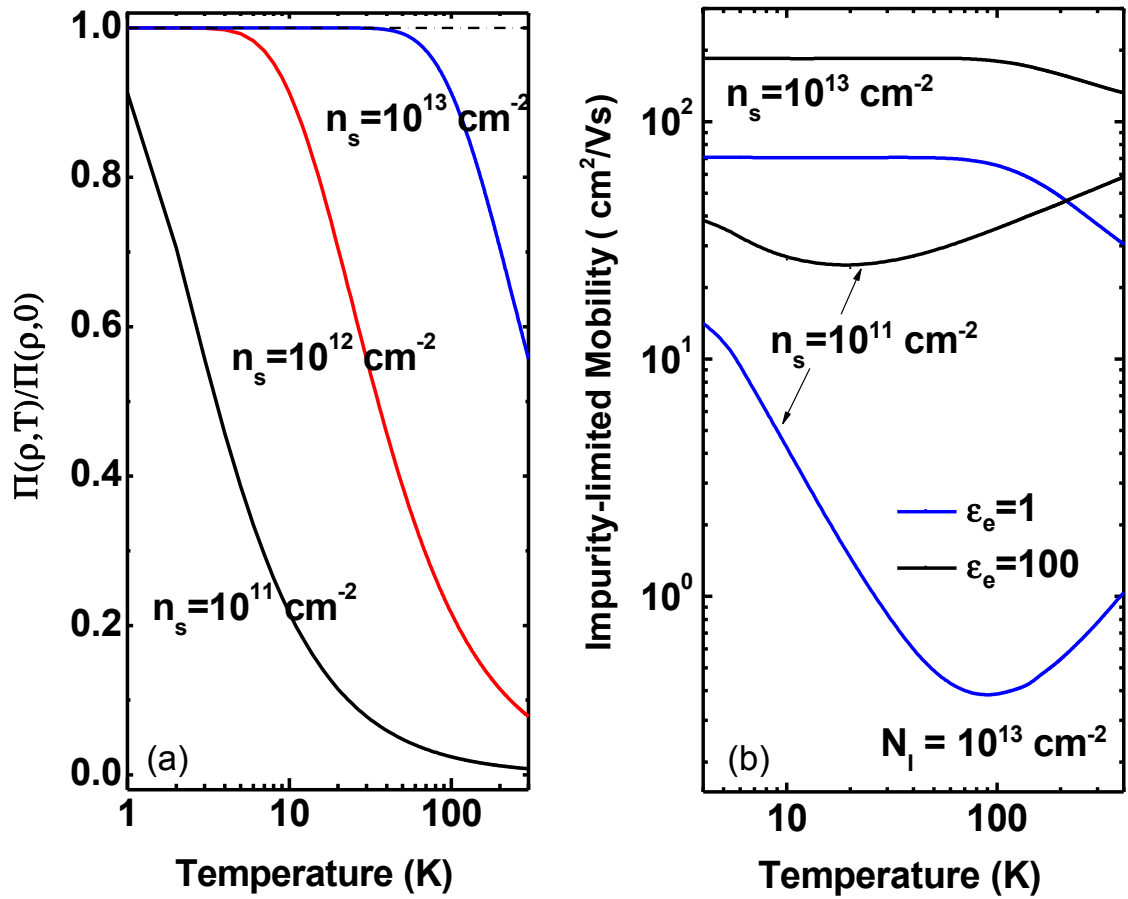

Figure 3

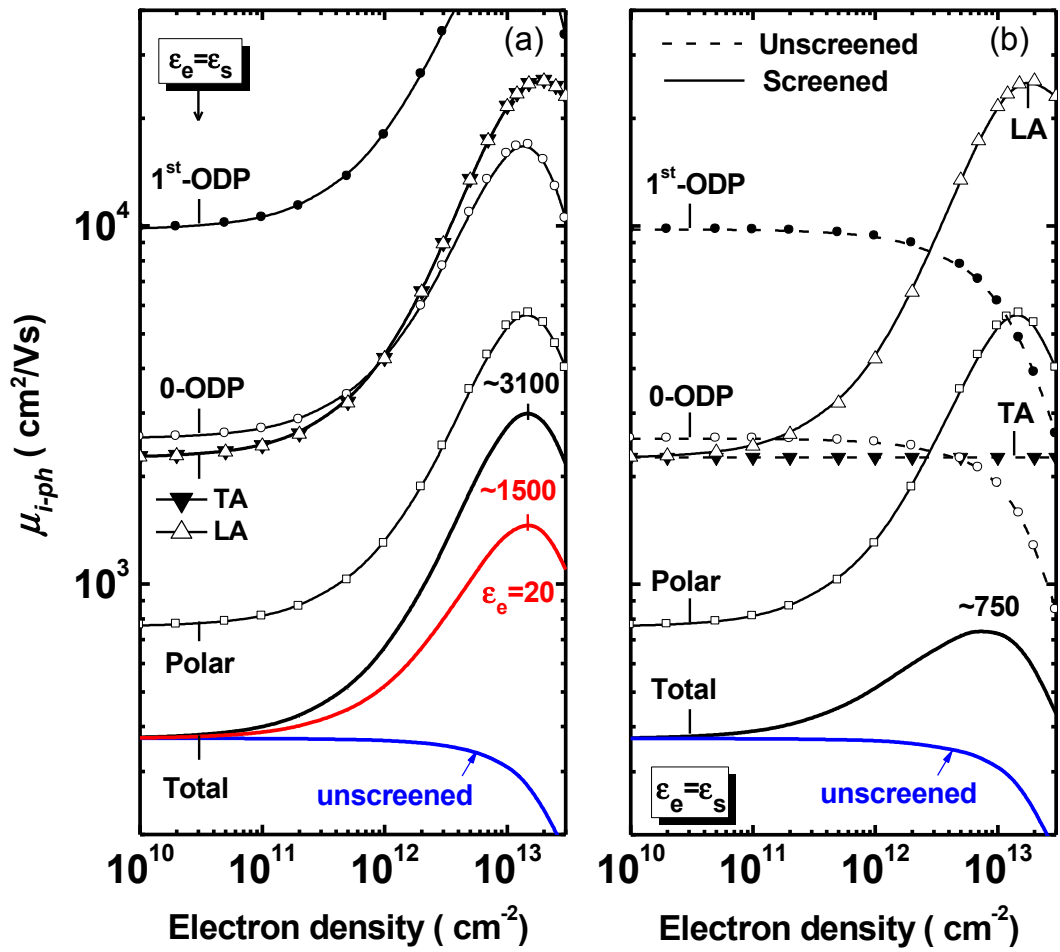

Figure 4

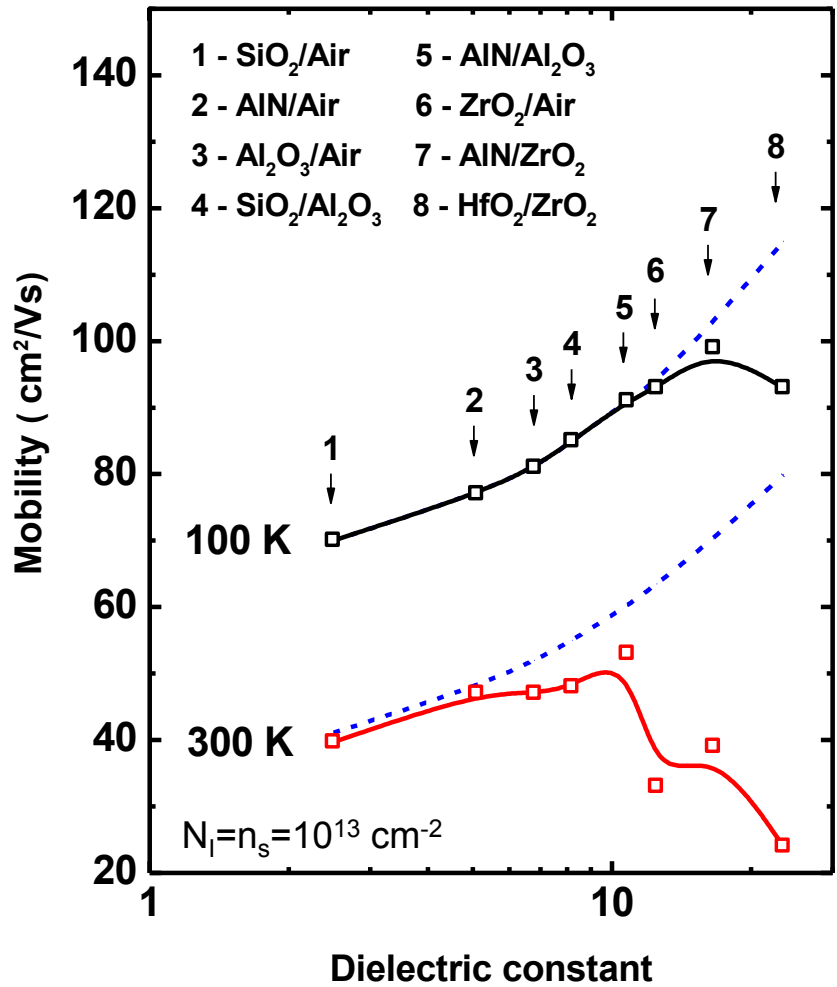

Figure 5

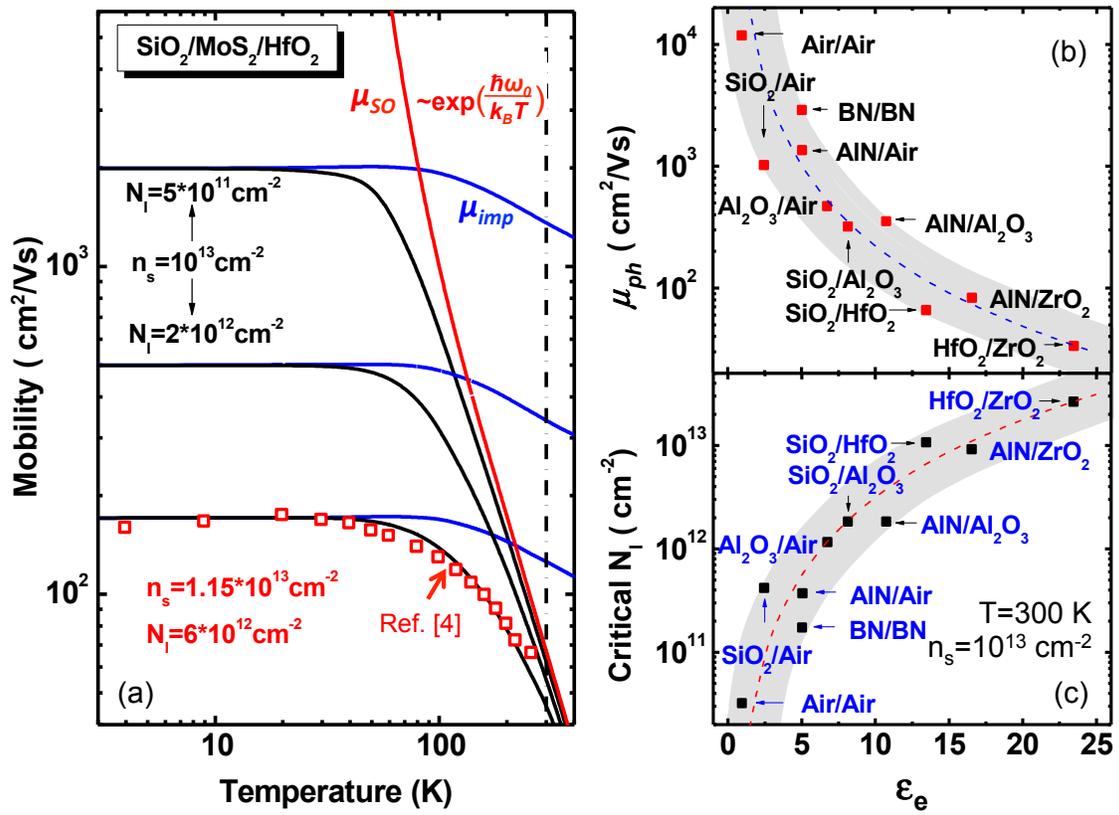

Figure 6

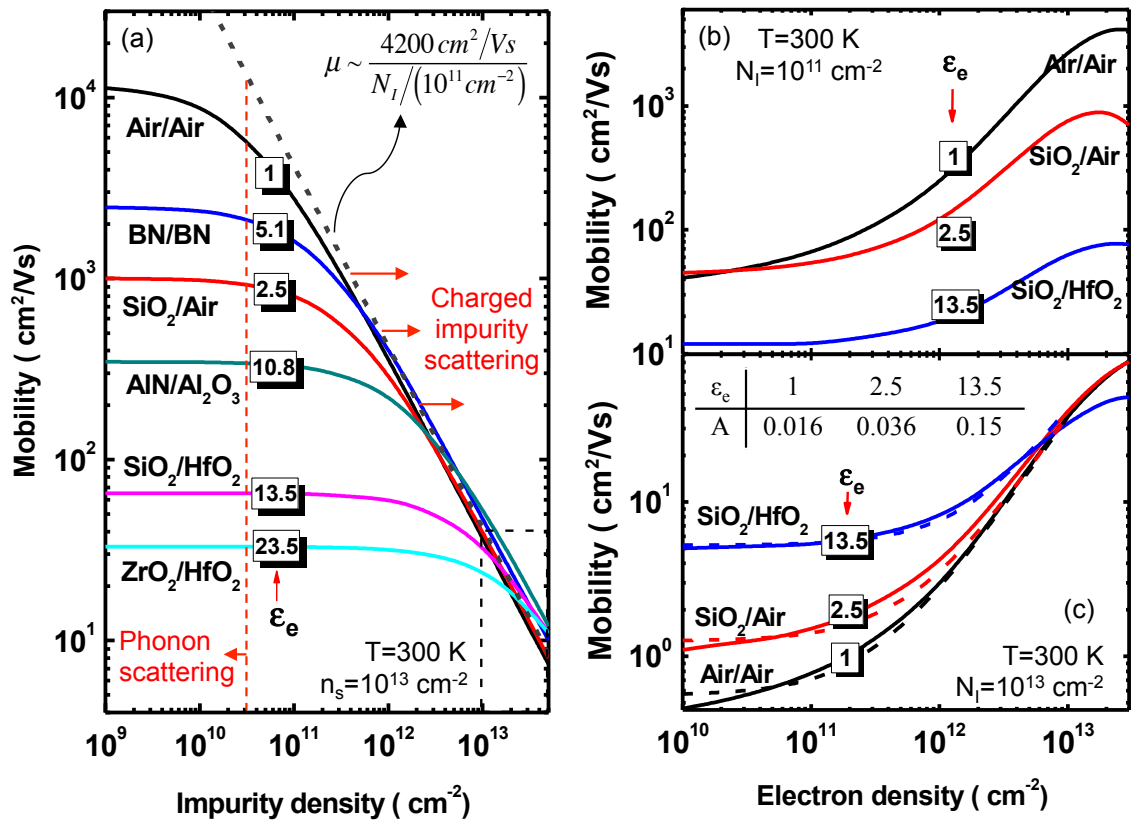

**Figure 7**